\documentstyle[11pt,Moriond,epsfig,wrapfig]{article}

\newcommand{\ee}{\mbox{$\mathrm{e}^{+}\mathrm{e}^{-}$}}

\def\be{\begin{equation}}
\def\ee{\end{equation}}
\def\bea{\begin{eqnarray}}
\def\eea{\end{eqnarray}}

\begin{document}
\unitlength 1cm
\vspace*{1.cm}
\title{HOW ROBUST IS THE RESULT OF THE STANDARD MODEL HIGGS BOSON SEARCH AT LEP?~\footnote{Talk given at the XXXVI$^{\scriptsize \hbox{th}}$ Rencontres de Moriond  Electroweak interactions and Unified Theories.}}

\author{Marumi Kado}

\address{European Organization for Nuclear Research (CERN),\\
1211 Geneva 23, Switzerland.}

\maketitle 

\abstracts{An excess of signal-like events above the expected
  background, corresponding to approximately three standard
  deviations, was observed in the search for the standard model Higgs
  boson at LEP in 2000. This excess is consistent with the existence
  of a Higgs boson with mass 115~GeV/c$^2$. Relevant consistency and
  robustness checks, which further support the signal interpretation,
  are presented.}  \vspace*{.5cm}

The outstanding run of LEP in 2000 at centre-of-mass energies up to
209~GeV allowed the standard model Higgs boson search sensitivity to
exceed three standard deviations ($\sigma$) up to a mass of
115~GeV/c$^2$. Amazingly enough, an excess of 2.9$\sigma$ was
observed~\cite{aleph} around this sensitivity limit.
Not only is this global excess in quantitative agreement with the
signal prediction, but its characteristics are also qualitatively
consistent with those expected from a 115~GeV/c$^2$ Higgs boson. More
explicitly, the distributions of the events among the
experiments~\cite{aleph,opal,delphi,l3}, among the decay
channels~\cite{aleph}, as a function of time, and as a function of
signal purity~\cite{aleph} all match the signal
hypothesis~\cite{request} with a global probability 33 times larger
than the background only hypothesis. Many consistency checks were
performed in order to exclude that this observation would result from
a systematic bias of the search. Two relevant examples are given here.  

\begin{figure}[h]
\begin{center}
\begin{tabular}{cc}
\mbox{\epsfxsize=.5\hsize\epsfbox{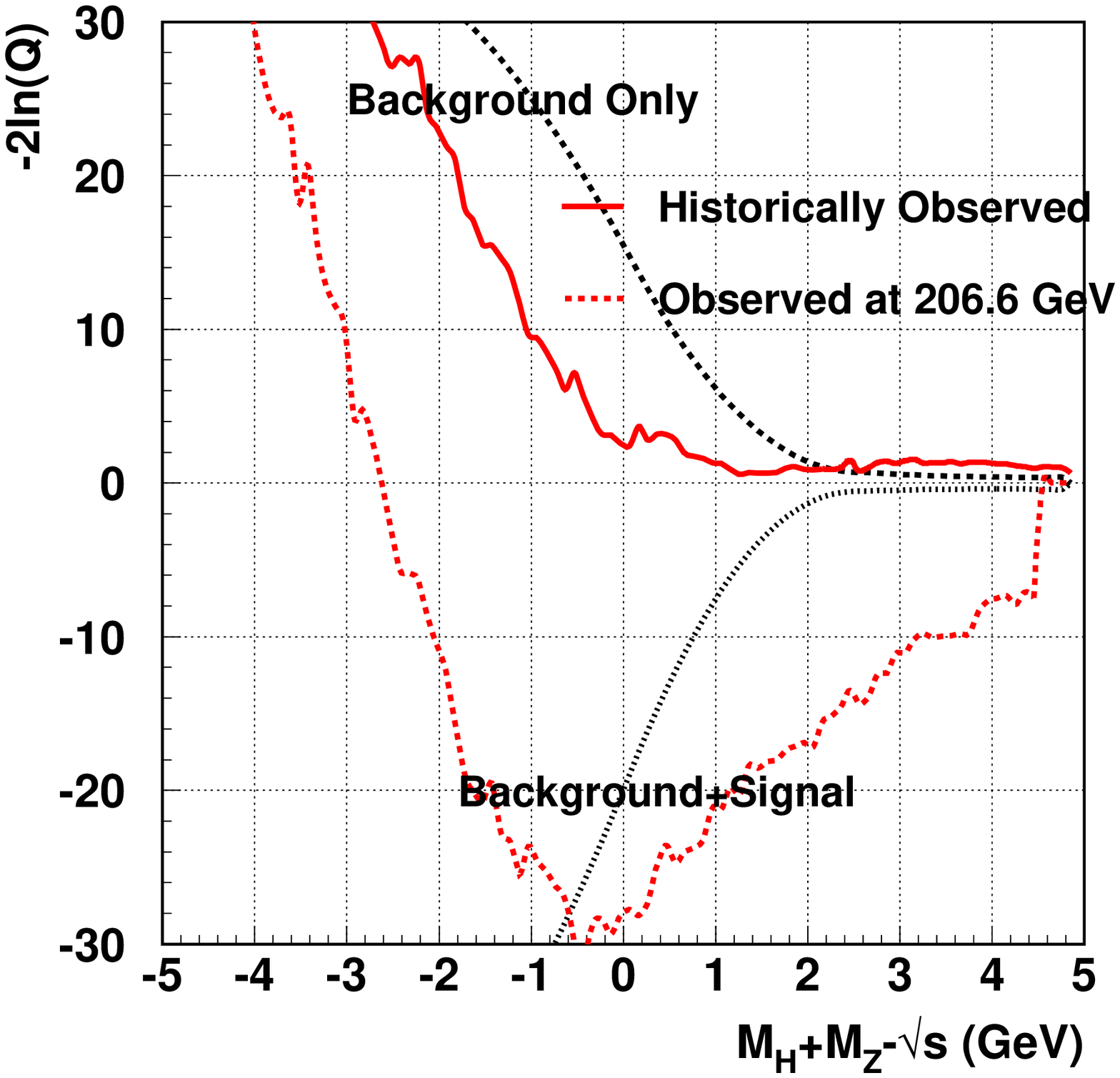}} 
\put(-1.35,1.5){(a)} &
\hskip-.5cm \mbox{\epsfxsize=.5\hsize\epsfbox{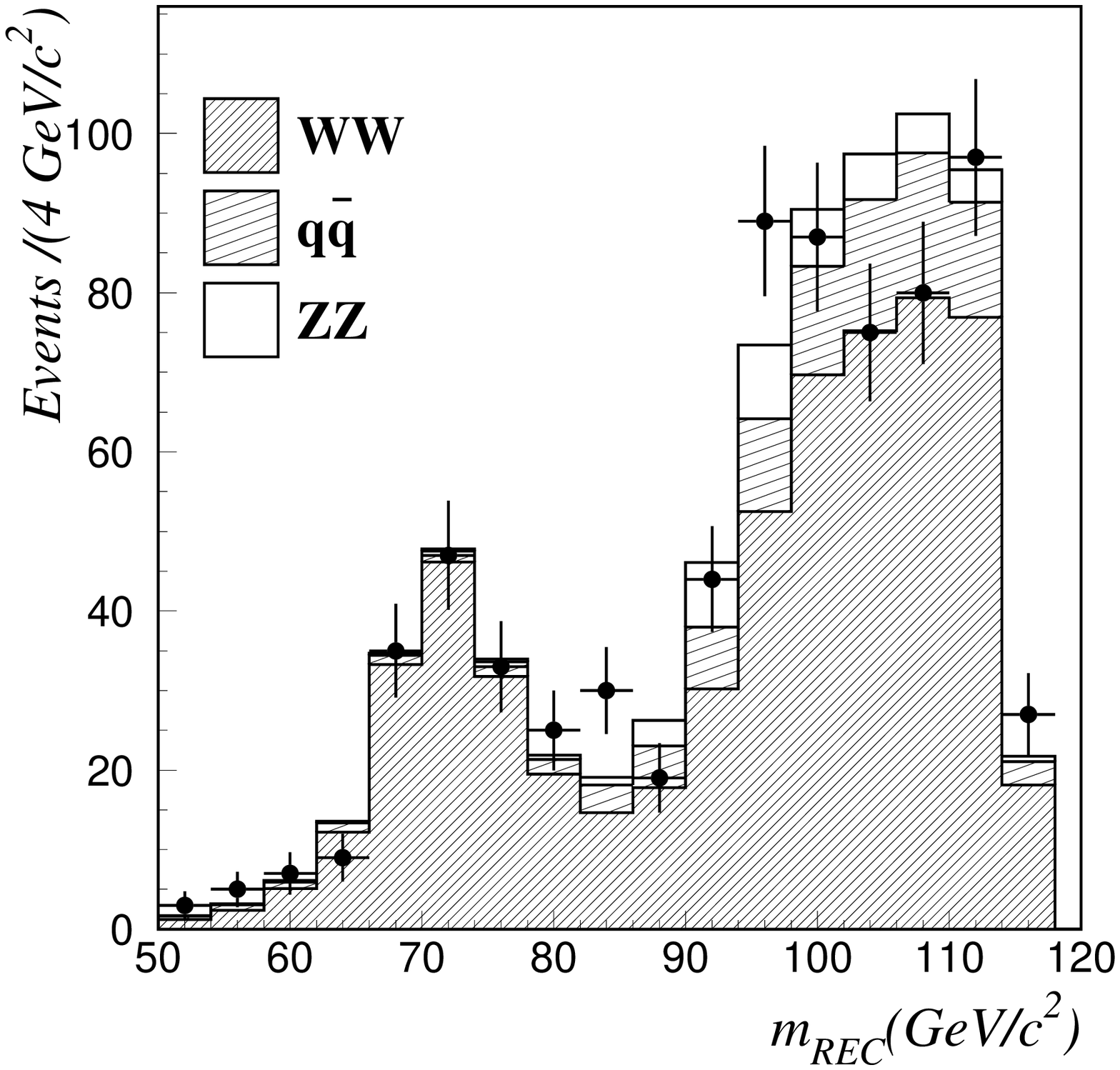}} 
\put(-6.5,4.4){(b)}
\end{tabular}
\vspace{-.5cm}
\caption{(a) 
  Observed estimators for the combination of all experiments and all
  data with $\sqrt{\rm s}<$206.5~GeV (full curve) and what it would
  have been had the excess observed above $\sqrt{\rm s}>$206.5~GeV
  been seen at all centre-of-mass energies (dashed curve) as a function of the
  distance of the Higgs boson mass hypothesis to the threshold. The
  expectation in absence and presence (dotted curves) of signal are
  also shown. (b) Distribution of the reconstructed mass for
  anti-b-tagged events selected by the ALEPH four jet ``Cuts'' stream
  analysis.
\label{llr}}
\end{center}
\end{figure}

In the first check, a combination of the searches with the
500~pb$^{-1}$ of data taken at centre-of-mass energies ranging from
189 to 206~GeV is compared to what it would have been had the excess
observed above 206~GeV been due to systematic bias close to the
kinematic threshold. The estimators, for both cases, are displayed in
Fig.~\ref{llr}-a~\cite{llrct} as a function of the distance to
threshold $(m_{\rm H}+m_{\rm Z}-\sqrt{s})$. This result largely
excludes the possibility of a systematic effect arising in the mass
reconstruction.

The second check is aimed at further illustrating the robustness of
the mass reconstruction algorithm in the four jets channel in
ALEPH~\cite{alephpaper}. A control sample is selected with the
kinematical and topological criteria of the ALEPH ``Cuts''
stream~\cite{aleph} analysis, but with the requirement that none of
the jets be tagged as b-quark jets~\cite{alephpaper}. This control
sample, mostly free of signal events due to the anti-b-tag criterion,
contains mainly WW events with kinematic similarities with the signal.
The reconstructed mass spectrum for the aforementioned control sample
is shown in Fig.~\ref{llr}-b.  The peak in the reconstructed mass
spectrum around $m_{\rm rec}\approx 2 \times m_{\rm W}-m_{\rm Z}$
corresponds to events where the jet pairing is correctly assigned to
each W boson.  The broader peak at higher reconstructed masses is due
to events with the wrong choice of pairing. The good agreement over
the full reconstructed mass spectrum
illustrates the robustness of the mass reconstruction procedure in the
experiment and the channel where the excess is most present (the fact
that most of the excess is seen in this particular channel is 
expected in the hypothesis of a signal).

In view of its consistency in all regards and its robustness, the
evidence for a signal is as strong as could be expected from the
amount of data collected at centre-of-mass energies above 206~GeV.
Besides, these hints are also consistent with the indirect constraints
on the mass of the standard model Higgs boson from precision
electroweak measurements~\cite{ew}.

\section*{Acknowledgements}
It is a pleasure to thank the organisers and the secretariat of the
XXXVI$^{\rm th}$ Rencontres de Moriond for their wonderful
hospitality. I am very gratful to Patrick Janot and Jean-Baptiste de
Vivie for their careful reading of these proceedings. I would like to
thank the European Community for financially supporting my
participation to this conference.


\section*{References}
\begin{thebibliography}{99}
  
\bibitem{aleph} P. Teixeira Dias, {\it ``Observation of an excess in
    the ALEPH search for the standard model Higgs boson, LEP combined
    results''}, these proceedings.
  
\bibitem{opal} I. Nakamura, {\it ``The Standard Model Higgs in
    OPAL''}, these proceedings.
  
\bibitem{delphi} P. Morettini, {\it ``Higgs searches in DELPHI''},
  these proceedings.

\bibitem{l3}
L3 Collaboration,
{\it ``Higgs candidates in e+ e- interactions at s**(1/2) = 206.6-GeV''},
Phys.\ Lett.\ B {\bf 495}, 18 (2000).
  
\bibitem{request} LEP Higgs Working Group, {\it Standard Model Higgs
    Boson at LEP: Results with the 2000 data, Request for running in
    2001}, http://alephwww.cern.ch/~janot/LEPCO/lephwg.ps
  
\bibitem{llrct} LEP Higgs Working Group, {\it FAQ (and their answers)}, \\
  http://alephwww.cern.ch/~janot/LEPCO/qanda.ps
  
\bibitem{alephpaper}
ALEPH Collaboration,
{\it ``Observation of an excess in the search for the standard model Higgs  boson at ALEPH''},
Phys.\ Lett.\ B {\bf 495}, 1 (2000).

\bibitem{ew} E. Tournefier, {\it ``Electroweak results and fit to the
    standard model''}, these proceedings.

\end{thebibliography}
\end{document}